# Light induced disassembly of dusty bodies in inner protoplanetary discs: implications for the formation of planets

Gerhard Wurm

Institut für Planetologie
Wilhelm-Klemm-Str. 10
D-48149 Münster, Germany
Tel. +49 251 83 39052
Fax. +49 251 83 36301

e-mail: gwurm@uni-muenster.de.

## Abstract

Laboratory experiments show that a solid state greenhouse effect in combination with thermophoresis can efficiently erode a dust bed in a low pressure gaseous environment. The surface of an illuminated, light absorbing dusty body is cooler than dust below the surface (solid state greenhouse effect). This temperature gradient leads to a directed momentum transfer between gas and dust particles and the dust particles are subject to a force towards the surface (thermophoresis). If the thermophoretic force is stronger than gravity and cohesion dust particles are ejected. Applied to protoplanetary discs, dusty bodies smaller than several km in size which are closer to a star than ~ 0.4 AU are subject to a rapid and complete disassembly to sub-mm size dust aggregates by this process. While an inward drifting dusty body is destroyed the generated dust is not lost for the disc by sublimation or subsequent accretion onto the star but can be reprocessed by photophoresis or radiation pressure. Planetesimals cannot originate through aggregation of dust inside the erosion zone. If objects larger than several km already exist they prevail and further grow by collecting dust from disassembled smaller bodies. The pile-up of solids in a confined inner region of the disc in general boosts the formation of planets. Erosion is possible in even strongly gas depleted inner regions as observed for TW Hya. Reprocessing of dust through light induced erosion offers one possible explanation for growth of large cores of gas poor giant planets in a gas-starved region as recently found around HD 149026b.

## 1. Introduction

Terrestrial planet formation and the formation of giant planet cores in accretion models require mechanisms to assemble the solid fraction, initially dust or ice particles, in protoplanetary discs (Weidenschilling 2000; Dominik et al. 2007; Nagasawa et al. 2007). Collisions between particles of all sizes are among the most basic processes to achieve growth. Collisions of particles produce dusty bodies which are larger than 10 cm (Dominik et al. 2007). These dust aggregates stick together by surface forces. Objects larger than km-size, called planetesimals, also grow through collisions (Wetherill & Stewart 1989; Kokubo and Ida 2002). Self-gravity aids reaccretion of fragments in collisions of these larger objects. Collisional growth to span the gap of m to km size is debated and is more complex. Collision velocities of up to several tens of m/s have to be considered (Weidenschilling 1977). It was shown in laboratory experiments that growth in high speed collisions at 25 m/s is possible (Wurm et al. 2005), but in less favourable situations of fragmentation, self gravity is not strong enough to confine fragments after a collision. Gravitational instabilities to form planetesimals are alternatively discussed (Goldreich & Ward 1973). The question of whether instabilities do form planetesimals in turbulent discs is subject to current research (Johansen et al. 2006, Rice et al. 2006).

Ideas of planetesimal formation highlight the constructive part of particle assemblage, but the destructive processes that disassemble large bodies play an important role in planet formation as well. Fragments can be redistributed within the disc (Ciesla & Cuzzi 2006; Cuzzi & Weidenschilling 2006). Fragmentation in one place can locally enhance solid particle densities in other places and favour planet formation. It is long known that m-size objects drift *inward* at a speed of 1 AU in 100 y in a typical disc (Weidenschilling 1977). These solids provide a potential reservoir of matter to be incorporated into larger bodies closer to a star. In gas depleted, transparent protoplanetary discs, particles smaller than 10 cm in size are transported *outward* by photophoresis. This leads to transport of material from the inner solar system to the comet forming region (Krauss & Wurm 2005; Mousis et al. 2007).

Collisions are the most prominent mechanisms to produce fragments. A recent work by Paraskov et al. (2006) finds that also pure gas flow can erode dusty planetesimals on eccentric orbits in inner protoplanetary discs. While this mechanism requires the presence of a dense gas, the gas density decreases with time due to viscous evolution, photo evaporation and planet formation (Najita et al. 2007; Alexander & Armitage 2007). Erosion by gas is not possible in a gas depleted disc, but at the same time the light of the central star can penetrate deeper into the disc.

We introduce here an erosion mechanism based on the star's illumination. Illuminated dusty bodies are subject to complete disassembly in the inner few tenths of AU from a star. This erosion effect has only recently been found in experiments (Wurm & Krauss 2006). In section 2 we describe the principles of light induced erosion. In section 3 we describe laboratory experiments. In section 4 we apply the experimental results to protoplanetary discs. Section 5 and 6 give some further discussion and conclusions.

## 2. Light induced erosion

If dark dust is illuminated by visible light, dust aggregates are ejected from the surface (Wurm & Krauss 2006). The occurrence of this effect depends on the gas pressure and light intensity. The erosion of a dusty body can be explained in terms of two effects, a solid state greenhouse effect and thermophoresis.

**Solid State Greenhouse Effect:** We consider a body in thermal equilibrium with its surroundings. The body is illuminated on one side by visible light. We first assume the classical case, that the body absorbs radiation only at the surface. The body heats up and a temperature gradient is established where the surface is always at the highest temperature. The temperature steadily decreases from the surface to layers deeper under the surface (Davidsson & Skorov 2002a).

Often radiation is not only absorbed at the surface but can penetrate below the surface. An example is ice with some absorbing inclusions below the surface (dirty ice). The body is heated in a thick top layer of the solid. Far below this top layer, where no more light is absorbed, heat is transported by heat conduction and again the classical temperature gradient is established. This means that far below the surface the temperature still steadily decreases with depth.

However, in this case it is important that the surface is cooled by radiation. It has to be noted that, as the solid is opaque to thermal (infrared) radiation, only the surface cools by radiation not the whole top layer. Due to this additional cooling of the surface and depending on the functional behaviour of light absorption with depth a temperature gradient can be established where the temperature steadily *increases* from the surface to deeper layers. As the temperature decreases with depth far below the top layers, this requires that the temperature reaches a maximum somewhere *below* the surface.

In planetary atmospheres a similar process occurs. Visible sunlight is transmitted through the atmosphere and is absorbed on the ground. Thermal radiation from the ground cannot pass the atmosphere without partial extinction. The ground layers of the atmosphere heat up with respect to the outer atmosphere. This is well known as greenhouse effect. In analogy to this but occurring in a solid body the temperature increase with depth below the surface of a solid body as outlined above has been termed solid state greenhouse effect (Niederdorfer 1933; Brown & Matson 1987; Kömle et al. 1990; Davidsson & Skorov 2002a; Kaufmann et al. 2006). The idea of the solid state greenhouse effect is sketched in Fig. 1. Quantitative temperature profiles with depth as described above qualitatively and indicated schematically in Fig. 1 can, e.g., be found in Davidsson & Skorov (2002a), Kaufmann et al. (2006), Kömle et al. (1990), or Wurm & Krauss (2006).

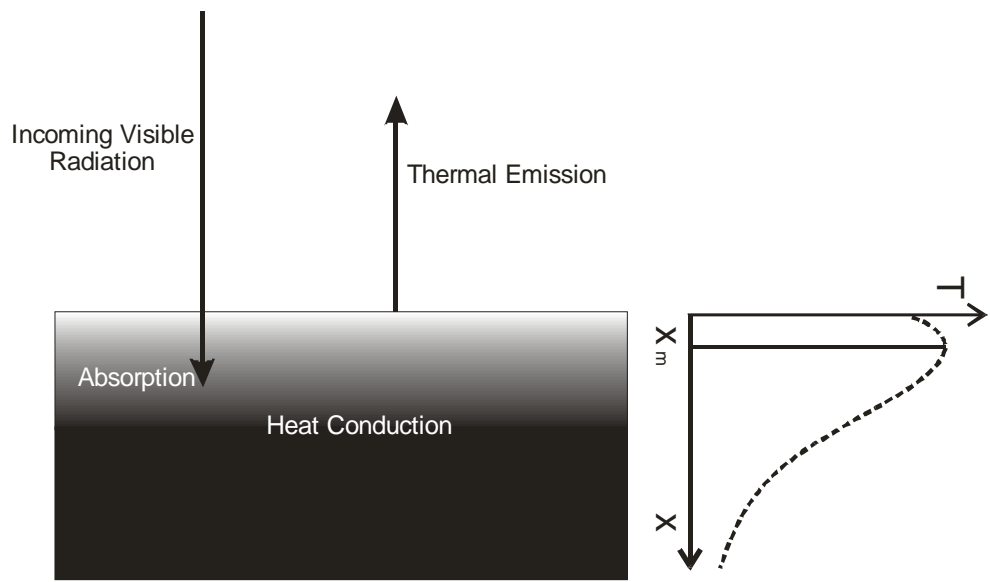

**Fig. 1 Solid state greenhouse effect. Incoming, directed visible light is absorbed and heats up the solid. Within the solid heat is transported by conduction. The surface cools due to thermal emission. A temperature gradient (right) develops with the maximum temperature being at a certain depth $x_m$ below the surface. For quantitative temperature profiles refer to Davidsson & Skurov (2002a), Wurm & Krauss (2006), Kömle et al. (1990), or Kaufmann et al. (2006).**

In this paper we do not consider solid, semi-transparent bodies with little or no porosity, but highly porous dust beds which consist of absorbing dust particles. Here, light is transferred deeper into the dust bed through the pores of the dust bed. In addition it is scattered to deeper layers by forward light scattering if the dust particles are larger than the wavelength. As a result, a solid state greenhouse effect occurs even in dust beds consisting of highly absorbing material like graphite as experimentally shown (Wurm & Krauss 2006).

**Thermophoresis:** If a particle is embedded in a gas it moves from the warm to the cold side in a temperature gradient. This motion is known as thermophoresis. At high gas pressure (continuum regime) the thermophoretic force is a reaction to thermal creep. Thermal creep is the gas motion at the surface of the particle from the cold to the warm side (Vedernikov et al. 2005). At low pressure (free molecular flow regime) the thermophoretic force results due to a difference in momentum transfer to or from gas molecules, as a fraction of molecules accommodate on the surface. They impinge the surface, stick for a short time, and are then released diffusely with a kinetic energy according to the local surface temperature. Gas molecules on the warm side on average leave faster than on the cold side. To balance the net momentum the particle has to move from the warm to the cold side.

An important feature of thermophoretic forces is the dependence on gas pressure. In the free molecular flow regime the force increases with increasing pressure. In the continuum regime the force decreases with increasing pressure. A semi-empirical equation for the thermophoretic force for all pressures if a linear temperature gradient $dT/dx$ across spherical particles of radius $a$ exists, is given by Rohatschek (1995)

$$F_{th} = F_{max} \frac{2}{\dfrac{p}{p_{max}} + \dfrac{p_{max}}{p}}$$

, (1)

$$p_{max} = D\sqrt{\frac{2}{\alpha}}\frac{3T}{\pi a}, \quad F_{max} = D\sqrt{\frac{\alpha}{2}}a^2\frac{dT}{dx}, \quad D = \frac{\pi}{2}\sqrt{\frac{\pi}{3}}\kappa\frac{c\eta}{T}, \quad c = \sqrt{\frac{8}{\pi}\frac{R_g T}{\mu}}$$

where, $F_{max}$ is the maximum force achievable at a pressure $p_{max}$. It is $\mu$ the molar mass of the gas, $R_g$ is the gas constant, and the dynamic viscosity of the gas is $\eta$. It is $\kappa$ the thermal creep parameter which is related to the thermal accommodation coefficient $\alpha$. Both are often close to 1.

In the top layers of a porous dust bed gas moves freely and a thermophoretic force acts on every dust particle with a temperature gradient. Due to the solid state greenhouse effect particles are dragged toward an illuminated surface. In static equilibrium, this is compensated by cohesion and gravity. If the thermophoretic force exceeds the gravitational force on a particle, movement of the particle is hindered only by cohesion. The drag on particles deeper under the surface is a superposition of all particle drag forces above this particle and cohesion has to balance a drag force increasing with depth. If the total drag force exceeds the maximum sticking force contacts break. An aggregate consisting of many dust particles is released. A sketch of this thermophoretic eruption principle is shown in Fig. 2.

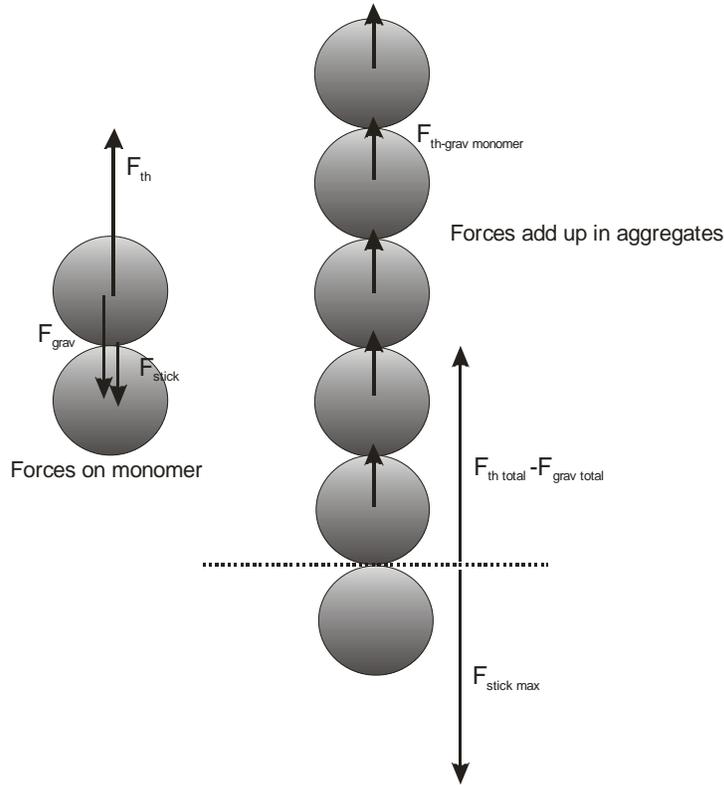

**Fig. 2** Thermophoretic forces $F_{th}$ and eruption in a temperature gradient; every particle is subject to a thermophoretic force, gravity, and a sticking force (left). In a static configuration the sticking force (surface force) adjusts as to balance the difference between thermophoretic force and gravity. For a chain of particles the forces add up. The result is an increasing upwards directed drag force balanced by an ever increasing sticking force at particle contacts. Eventually, the lifting force reaches the maximum of the sticking force, and the contact breaks at the dotted line (right). The chain relaxes and feels the full superposition of thermophoretic (decreased by gravitational) forces accelerating it – an aggregate is ejected.

If the drag force at the depth of maximum temperature is not sufficient to break contacts established by surface forces, the illuminated dust bed is stable. Ejection of particles can still occur and is observed when the light source is turned off. Without illumination the temperature gradients decrease within the dust bed but heat transfer calculations show that the position of the temperature maximum $x_m$ moves deeper down below the surface at the same time (Wurm & Krauss 2006). While the thermophoretic force on an individual particle decreases with decreasing temperature gradient, the spatial extent of the top layer susceptible to thermophoresis increases. The total thermophoretic force added up for many particles can be stronger than before and leads to an eruption as the light source is turned off. In the context of bodies in protoplanetary discs this is important if bodies rotate and if a surface moves from the illuminated side into the shadow or if a surface is shadowed otherwise.

3.   Experiments on light induced erosion

Fig. 3 shows a sketch of the experimental setup which has been used to study ejection of particles due to thermophoresis and a solid state greenhouse effect. A dark dust powder is sieved into a tray through a 0.5 mm mesh. This dust bed is placed into a vacuum chamber. A laser is directed onto the dust surface 23° (± 3°) inclined towards the vertical and the dust, i.e. dust ejection, is observed through a stereo microscope at varying gas pressure. A sequence of images which shows an eruption of dust is seen in Fig. 4.

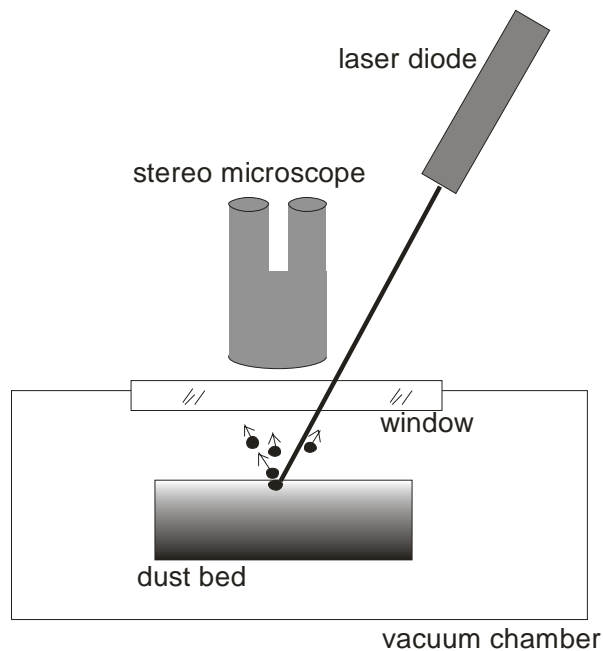

**Fig. 3.** Sketch of the experimental setup

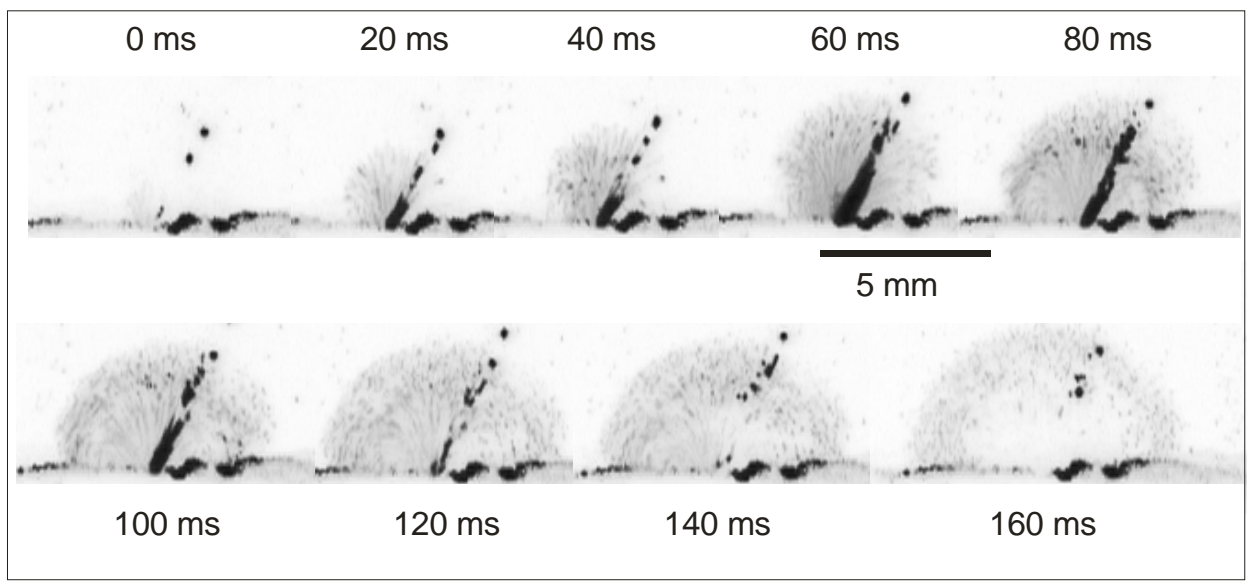

**Fig. 4. Eruption of vitreous carbon dust observed with a digital video camera; from top left to bottom right still frames spaced in 20 ms intervals are shown. The colour images have been converted to grey scale and inverted for visibility. Dust shows up as grey or black. The images are a side view of the dust bed. The thick diagonal line is the location of the laser beam which leads to eruptions (comp. fig. 3). Ejected dust dominantly shows here as it is illuminated by the laser. To visualize the total extent of the eruption a second, weaker laser provides a light curtain in the focal plane. It illuminates a thin section of the fragment cloud which spreads from the surface. The gas pressure is ~ 150 mbar and laser intensity is about 100 kW m$^{-2}$ in this case. Note that the material (vitreous carbon) and intensity in the slightly adjusted setup to take these images deviates**

**from the material (graphite) and intensity of 50 kW m$^{-2}$ used to determine the erosion threshold used in this paper otherwise.**

In the experiments a laser diode of about 50 mW as light source was used. The spot size was approximately 1.6 mm x 0.7 mm in experiments for which the pressure dependence of erosion was quantified as given below. From this we estimate the light intensity to be $I = 50$ kW/m$^2$.

At the given intensity we observe ejection for all tested dark dust powders at low (~ mbar) pressure, i.e. we observed dust eruptions for graphite, basalt, vitreous carbon, and carborundum. Dust particles within these powders were up to 100 μm in size but the size distribution was not quantified in detail. Continuous erosion could be obtained for these dust samples in a certain pressure range. To quantify the process we determined the lowest pressure $p_{th}$ at which a dust bed which is moved through the laser can continuously be eroded to deeper layers. The basalt sample has the highest threshold pressure of $p_{th} = 1$ mbar ($\pm$ 0.5 mbar) and is therefore stable at lower pressure while the other dust samples still get eroded. It is therefore the worst case with respective to erosion. Therefore, we only use the basalt values for quantitative calculations further on. In earlier experiments we quantified the pressure dependence of the ejection mechanism in more detail (Wurm & Krauss 2006). We found that the thermophoretic force shows the characteristic incline and decline with pressure in agreement to equation (1). We assume that the temperature gradient is proportional to the light intensity. The thermophoretic force according to equation (1) in the free molecular flow regime is thus given as

$$F_{th} = b \frac{Ip}{T},$$ (2)

where all properties of the dust bed with respect to thermophoresis are put into parameter $b$. The other parameters $I, p,$ and $T$ depend on the environment of the particle. In the laboratory experiments we find erosion for basalt if the thermophoretic force is larger than

$$F_{th\_e} = b \frac{I_e p_e}{T_e} = b \cdot 15 \text{ kWm}^{-2} \text{ Pa K}^{-1},$$ (3)

based on $p_e = 1$ mbar, $I_e = 50$ kW/m$^2$, and an ambient temperature $T_e = 300$ K. By dividing equation (2) by equation (3) we define a dimensionless erosion parameter $\chi$ for a given dust bed

$$\chi = \frac{\frac{Ip}{T}}{15 \text{ kWm}^{-2} \text{ Pa K}^{-1}}.$$ (4)

Continuous erosion occurs for $\chi \geq 1$.

According to the model of ejection given above, eruptions occur after thermophoretic forces are summed up over a larger number of dust grains. The calculated typical size of an actively ejected aggregate is about 100 μm (Wurm & Krauss 2006). As soon as the erosion threshold is crossed a laser beam produces mm deep craters within fractions of a second.

4. Application to protoplanetary discs

4.1. Erosion zones

Protoplanetary discs are thought to be truncated at some inner edge. It is important to note that two inner edges exist – one for gas and one for dust. Eisner et al. (2005) find observationally that the dust toward several T-Tauri stars is truncated between 0.1-0.3 AU from the star. Optically thin gas might be present further inwards. In a few cases there is a clear distinction between inner edges of gas and dust (Najita et al. 2007). So called transitional discs have rather large inner dust free regions (Sicilia-Aguilar et al. 2006). In the case of TW Hya the inner dust depleted region is 4 AU in radius (Calvet et al. 2002). This inner region is supposed to be optically thin in the visible in radial direction. Still, dust is present at 0.06 AU (Eisner et al. 2006). Though comparably tenuous, a dynamically significant amount of gas is also still present in TW Hya at about 0.2 AU (Salyk et al. 2007). Salyk et al. (2007) give an estimate for the surface density of gas in the inner 1 AU of TW Hya of 1 g cm$^{-2}$.

Alexander and Armitage (2007) model the clearing of protoplanetary discs by photo evaporation and associated motion of dust. They nicely demonstrate that in a certain phase dust particles are embedded in gas *and* are illuminated by the star. They do not consider photophoresis, but their calculations give a direct indication of the evolution of protoplanetary discs through phases where light induced erosion and photophoresis are important.

To summarize, there is little doubt that protoplanetary discs – over some period of their life – consist of an inner disc which is gaseous and transparent outside of the dust sublimation radius. Erosion by a solid state greenhouse effect and thermophoresis occur in these inner regions if the threshold with respect to gas pressure, light intensity, and temperature as given in equation (4) is reached or if $\chi \geq 1$. A light induced erosion zone extents from the inner truncation radius of the gas disc to the distance where the erosion parameter $\chi(R) = 1$.

The extent of the erosion zone depends on the inner disc model. Many competing models exist, including viscous evolution, photo evaporation, disc-planet interactions, dust-gas coupling. At this stage of introducing the new erosion model to protoplanetary discs we choose a power law dependence of the gas density and pressure. To quantify possible erosion zones we adopt a disc profile in analogy to Hayashi et al. (1985) as starting point.

$$\rho = f \cdot 1.4 \cdot 10^{-6} \left(R/1\mathrm{AU}\right)^{-\frac{11}{4}} \mathrm{kg\,m}^{-3}$$

$$p = f \cdot 0.5 \cdot 10^{-2} T \left(R/1\mathrm{AU}\right)^{-\frac{11}{4}} \mathrm{Pa}$$

$$\Sigma = f \cdot 1.7 \cdot 10^3 \left(\frac{R}{1\,\mathrm{AU}}\right)^{-\frac{3}{2}} \mathrm{g\,cm}^{-2} \tag{5}$$

$$T = 280 \left(R/1\mathrm{AU}\right)^{-\frac{1}{2}} \mathrm{K}$$

It is $\Sigma$ the surface density of the disc, $R$ is the distance to the star. Gas mass density $\rho$, pressure $p$, and temperature $T$ are for the midplane of the disc. We introduced a density factor $f$ to cover different gas mass loading of the inner region. This might be due to different disc profiles (Alibert et al. 2005) or due to viscous evolution or photo evaporation over time.

To calculate the erosion parameter $\chi$ (equation 4) the light intensity has to be specified which is

$$I(R) = I_0(R) e^{-\sigma \int_{R_i}^{R} \rho dx}, \tag{6}$$

where $R_i$ is the inner truncation radius of the gas disc and $I_0$ is the intensity of the stellar radiation without extinction. The exponential function accounts for radial extinction within the disc. We only consider extinction due to Rayleigh scattering by molecular hydrogen which – in the absence of a significant amount of dust – is a dominant opacity source for the densities and temperatures considered (Mayer and Duschl 2005). It is σ the cross section for Rayleigh scattering by molecular hydrogen. We use a Rayleigh scattering cross section integrated over a blackbody spectrum of 6000 K of $\sigma = 5 \cdot 10^{-4}$ cm$^2$ g$^{-1}$ (Dalgarno and Williams 1962, Vardya 1962, Mousis et al. 2007).

Using equation (5) the intensity (equation 6) can be calculated analytically as

$$\int_{R_i}^{R} \rho dx = f \, 1.2 \cdot 10^5 \left( \left( \frac{R_i}{1\text{AU}} \right)^{-\frac{7}{4}} - \left( \frac{R}{1\text{AU}} \right)^{-\frac{7}{4}} \right) \text{kgm}^{-2} . \qquad (7)$$

In the following we consider a star with solar luminosity ($I_0 = 1367$ W/m$^2$ at 1 AU). We vary two parameters; the inner radius of the gas disc $R_i$ and the density factor $f$. Fig. 5 shows the width of the erosion zone for different inner hole sizes (radii $R_i$) in the gas disc depending on the density factor $f$.

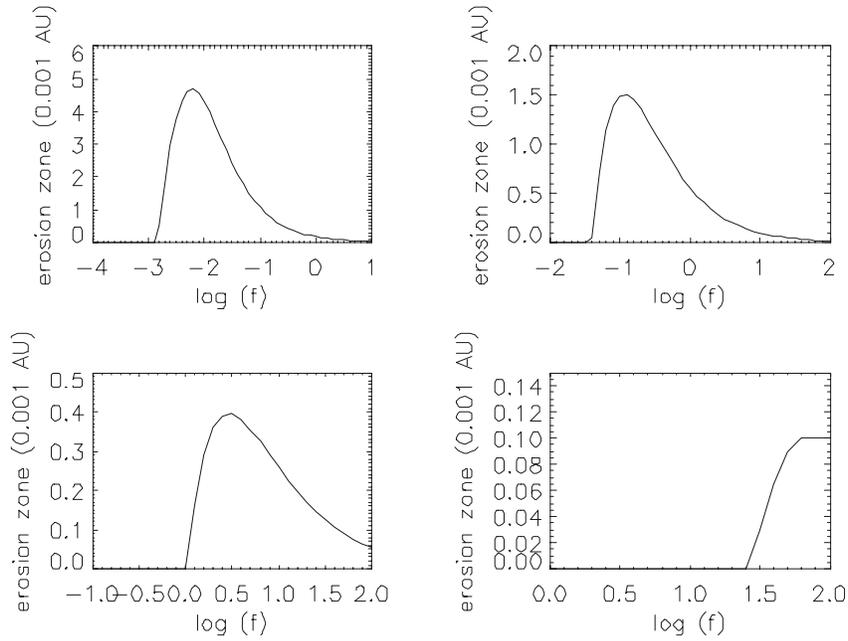

**Fig. 5 Erosion zones; depending on the size of the inner hole the maximum width of the erosion zone is given. From top left to bottom right the inner hole size is 0.05 AU, 0.1 AU, 0.2 AU, 0.4 AU. The density factor (x-axis) is multiplied to the profile given by Hayashi et al. (1985) (equation 5) and accounts for less dense or denser inner discs.**

Three distinct features can be seen for a given inner radius.

- For high gas densities the erosion boundary has little extent and essentially traces the inner edge of the gas disc. This is due to the large extinction by Rayleigh scattering.
- For very low gas densities, the erosion threshold is no longer reached. Note though that the density has to be decreased by a factor of $10^{-3}$ to rule out erosion at 0.05 AU.
- For intermediate densities erosion occurs in a wider zone from the inner edge to a certain distance, with a maximum at densities where extinction is not too high yet at some distance into the disc but where the gas pressure is still high enough to promote erosion.

As the inner radius $R_i$ is increased the erosion zone moves outwards and gets smaller. The maximum extension of the erosion zone is about 0.01 AU for small gas free regions. Erosion can reach to about 0.4 AU from the star (i.e. the orbit of Mercury) if we allow for a more massive disc.

### 4.2.    Size limit of erosion

Based on the experiments, we estimate that a body which is illuminated on the whole surface ejects more than a 1 mm thick layer per second. If these ejecta escape, further erosion is possible and a 1 km object is eroded on a time scale of $10^6$ s which is much less than 1 year. Ejecta of 100 μm in size couple well to the gas and initially stay close to the surface. In a sub-Keplerian disc a large body on a Keplerian orbit is embedded in a gas flow and the ejected particles get entrained in this gas flow which removes the ejecta. Only if gas drag is not stronger than the gravitational force of the eroded body does the body hold on to its ejecta. The drag force on an ejected particle of mass $m$ in a gas flow of velocity $v$ with a gas-grain coupling time $\tau$ is given by

$$F_{drag} = \frac{mv}{\tau} . \tag{8}$$

The particle is pulled back by gravity with

$$F_g = G \frac{mM}{S^2} = Gm \frac{4\pi}{3} S \rho_b , \tag{9}$$

where $M$ is the mass of the body and $\rho_b$ is its density. It is $S$ the size of the body and $G$ is the gravitation constant. It follows that gas drag and gravity are comparable for an object of the threshold size $S = S_{thr}$

$$S_{thr} = \frac{3}{4\pi} \frac{v}{G \rho_b \tau} . \tag{10}$$

For a gas velocity in the boundary layer of v = 50mm/s, assumed to be 1 per mil of the velocity typical for large bodies in protoplanetary discs (Weidenschilling 1977) and a bulk density of $\rho_b$ = 1000 kg/m³ we get $S_{thr}$ = 180 km s / $\tau$. The gas-grain coupling time strongly depends on the gas pressure. If erosion in protoplanetary discs occurs at 1 Pa pressure $\tau$ for a 100 μm particle is on the order of 10 seconds (Blum et al. 1996). It follows that erosion is possible on bodies as large as 10 km.

### 4.3.    Light induced recycling

So far we calculated the possible location and width of the erosion zone, considering Rayleigh scattering by hydrogen only. With an optical thick dust disc, erosion only works inside of the inner dust edge and two distinct scenarios are possible. 1. The inner dust edge is within the maximum extent of the erosion zone calculated before. In this case, the dust edge is the outer boundary of the erosion zone. 2. The inner dust edge is outside of the erosion zone. Both situations are visualized in fig. 6.

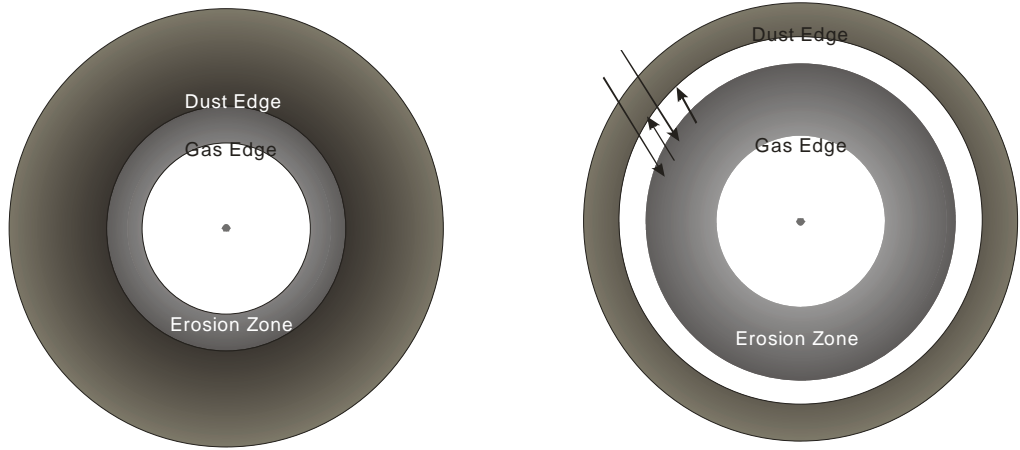

**Fig. 6.** Two possible scenarios in inner protoplanetary discs are shown. Left: The dust edge terminates the erosion zone. The dust edge is then continuously processed. Right: The inner edge of the dust disc is outside of the erosion zone. Recycling includes longer range transport. Large bodies move into the erosion zone, are disassembled, and the sub-mm dust aggregates are pushed back to the inner dust edge by photophoresis or radiation pressure in the case of small aggregates.

In the first case continuous erosion occurs at the dust edge. Bodies smaller than $S_{thr}$ consisting of dust cannot emerge from the edge and move further inward. Large bodies cannot grow at the edge from scratch as they are instantly destroyed by erosion again. Ejected particles can hardly get away from the scene of erosion. They are hindered of moving inwards as sub-mm particles move rapidly outwards if illuminated by light due to photophoresis (Krauss & Wurm 2005). In fact the forefront of the edge itself is pushed outwards as a whole by photophoresis (Krauss et al. 2007). As a result a very strong particle concentration of sub-mm dust aggregates occurs at the dust edge. If planetesimals larger than $S_{thr}$ are present at the dust edge they grow further by collecting the debris of smaller bodies which are eroded. This mechanism is restricted to within a few tenths of AU and works as close as the sublimation distance. The effect will influence the formation of planets in the proximity of the star. The formation of Mercury might have been influence by accumulation of solids in an erosion zone of the solar system.

In the second case, if the inner edge of a protoplanetary disc is outside of the erosion zone particle concentration at the dust edge also occurs. Photophoresis transports particles smaller than a few cm generated in the erosion zone outwards (Krauss et al. 2007). The difference to scenario 1 is that particles are not continuously recycled by erosion at the dust edge any longer. Particles collide and grow to dust aggregates at least to sizes of several tens of cm (Dominik et al. 2007). Larger dust aggregates are no longer supported by photophoresis and will drift inwards if the disc profile still has retained its original pressure gradient (Weidenschilling 1977). These bodies cross the erosion distance and are eroded again. The material is then fed back to the edge of the dust disc by photophoresis and the whole process of particle growth can start over again. Material is also recycled in this case and the total solid density in the inner region of the disc increases.

Particles are not only pushed outward by photophoresis but radiation pressure can also transport particles efficiently. This is especially true for the small (1 μm) dust particles where radiation pressure can be larger than the star's gravity. Also larger (10 μm to 100 μm) fluffy aggregates can be pushed outwards as it has been shown numerically and experimentally that the ratio between radiation pressure force and gravity can remain as high as for the individual dust grains (Saija et al. 2003).

In the region between the erosion zone and the dust edge a variety of sizes can persist. This changes the details in the evolution of solids. A detailed treatment is beyond the scope of this paper but it is likely that the general concentration of solids in the inner region will boost the formation of large rocky planets. In scenario 2 the boost of planet formation is not limited to the inner few tenths of an AU. It proceeds further out as the dust edge migrates to larger distances.

5. Discussion

Salyk et al (2007) estimate a lower limit of the surface density of gas in TW Hya at 0.2 AU from CO emission to be $\Sigma = 1$ g cm$^{-2}$. This corresponds to a density factor of $f = 5 \ 10^{-5}$ (equation 5). Eisner et al. 2006 observed dust in TW Hya at 0.06 AU. It is likely that the tenuous gaseous disc extents to this close distance as well. With the given density reduction factor erosion in TW Hya in the innermost region is possible under favourable conditions despite the low gas density. Any large body up to a few hundred meters in size which consists of dust and is brought to an inner location e.g. by scattering by a planet further out can be subject to significant erosion. Krauss & Wurm (2005) give an estimate that for gas pressures $p > 4 \ 10^{-3}$ Pa photophoresis is stronger than the sun's gravity. Eroded matter of sub-mm size in TW Hya is rapidly pushed outward by photophoresis to the current inner dust edge at 4 AU. This enhances the solid density at the inner dust edge and provides material for boosting planet formation.

Alternatively, a planet inside of 4 AU can intercept the dust and grow solids in an otherwise gas depleted region. We suggest that such a mechanism is important for the formation of super Earth planets with several Earth masses (Rivera et al. 2005, Valencia et al. 2007). Sato et al. (2005) discovered a planet around HD 149026b with a core of about 70 Earth masses, a substantial fraction of its total mass. In a typical disc a core of this size gathers much more mass in gas. One way to explain large rocky planets is by accretion of solids in a gas-starved environment (Charbonneau et al. 2007). The erosion and recycle mechanism discussed here is able to explain the growth of a large rocky core within a thin gas. As the gas mass is depleted during the phase of light induced recycling, an existing protoplanet selectively grows by solids rather than by accretion of gas.

It is a long standing problem that meter-size bodies drift in rapidly (Weidenschilling 1977). It has often been assumed that these objects are accreted onto the star. Light induced recycling prevents this. While a meter-size dusty body is not safe from destruction, the disassembled matter remains within the disc.

If large solid dusty bodies are continuously drifting to the inner region and are recycled by light induced erosion and photophoresis, in time the density of solids reaches values higher than the gas density. Drift and collision velocities are then reduced and planetesimals form through collisions in the shadow of the dust edge.

Light induced recycling depends on a dusty constitution of the evolving large bodies. If a body is lithified, e.g. as fragment of a differentiated body, the mechanism does not work. Such objects survive in the erosion zone. The effect of gravity on the threshold for eruption is currently unknown but the effect is supposed to be favoured under reduced or missing gravity. Detailed dependence of erosion, e.g. on dust particle size, dust bed porosity, pore size, optical properties of the material, beam size, wavelength, temperature, gas flow, and rotation of the body still have to be addressed in the future. It is clear from the principle sketch of the mechanism of erosion outlined above that it depends on all these parameters. Calculations by Davidsson & Skorov (2002a, 2002b) with respect to cometary surfaces indicate that there is no simple analytical dependency of the solid state greenhouse effect on each of these parameters. In this paper we used dust beds which typically have porosities larger than 65%, where porosity is defined as fraction of void volume to total dust bed volume. There is little doubt that the absolute values for erosion zones will vary if parameters are varied but this will not change the principle results. The actual values for the extension of the erosion zone and erosion rates will change but the principle idea will hold.

6. Conclusion

Based on laboratory experiments we introduced a new dust dispersion mechanism. If a dust bed is illuminated a solid state greenhouse effect and thermophoresis lead to efficient erosion. This effect shows potential for a significant influence in the processing of dusty solid matter in inner protoplanetary discs. The mechanism is capable of disassembling large dusty bodies up to several km in size completely into sub-mm dust aggregates in inner parts of protoplanetary discs. Depending on the detailed disc model erosion occurs in the inner 0.4 AU of a disc for a solar type star.

The dust aggregates generated by erosion are subject to photophoresis or radiation pressure and the solid matter remains within the disc and is not lost to the star. Erosion of large objects which drift inward increases the solid to gas mass ratio. It provides a reservoir of sub-mm particles which can be built into next generations of larger bodies at the dust edge or boost the formation of a few very large rocky planets. One way or the other the destruction of protoplanetary bodies by light induced erosion might support the formation of large planetary bodies.

Acknowledgement

This work is funded by the Deutsche Forschungsgemeinschaft. I appreciate the very helpful comments of an anonymous reviewer.